# HYBRID HEURISTIC-BASED ARTIFICIAL IMMUNE SYSTEM FOR TASK SCHEDULING

Masoomeh sanei[1] and Nasrollah Moghaddam Charkari[2]

[1]Department of Electrical and Computer Engineering, Tarbiat Modares University, Iran, Tehran.
Masoomeh.sanei@modares.ac.ir
[21]Department of Electrical and Computer Engineering of Tarbiat Modares University, Iran, Tehran.
charkari@modares.ac.ir

***ABSTARCT***

*Task scheduling problem in heterogeneous systems is the process of allocating tasks of an application to heterogeneous processors interconnected by high-speed networks, so that minimizing the finishing time of application as much as possible. Tasks are processing units of application and have precedence-constrained, communication and also, are presented by Directed Acyclic Graphs (DAGs). Evolutionary algorithms are well suited for solving task scheduling problem in heterogeneous environment. In this paper, we propose a hybrid heuristic-based Artificial Immune System (AIS) algorithm for solving the scheduling problem. In this regard, AIS with some heuristics and Single Neighbourhood Search (SNS) technique are hybridized. Clonning and immune-remove operators of AIS provide diversity, while heuristics and SNS provide convergence of algorithm into good solutions, that is balancing between exploration and exploitation. We have compared our method with some state-of-the art algorithms. The results of the experiments show the validity and efficiency of our method.*

## 1. INTRODUCTION

With increasing the complexity of signal, image and control processing algorithms in embedded applications, high computational power to satisfy real-time constraints is vital. This can be achieved by parallel multiprocessors which are often heterogeneous in embedded and Distributed Computing Systems (DCS) [1].

The performance of a parallel application on distributed system is highly dependent on both the application characteristics, means the execution cost and data communication costs between tasks etc., and the platform features that are computational capacities of the processors, the number of processors, interprocessor communication bandwidth, memory size etc. To effectively exploit distributed systems, an important challenge is how to map some tasks to processors in order to achieve some objectives, such as load balancing, minimization of interprocessor communication, battery saving or some combination of them for computationally demanding tasks with diverse computing needs. The widespread use of distributed computers in many computational-intensive applications makes the problem of mapping programs in distributed computers more crucial. The task assignment (or mapping) problem, also called as task scheduling problem, is the tasks assignment of an application to different processors in a distributed computer system in order to reduce the program turnaround time (or makespan) and to increase the system throughout [2-5].

Reducing the makespan leads to load balancing and minimization of interprocessor communication automatically. The scheduling problem is NP-complete for most of its variants except for a few simplified cases. These cases are (1) scheduling tree-structured task graphs with uniform computational costs on an arbitrary number of processors, (2) scheduling arbitrary task graphs with uniform computational costs on two processors and (3) scheduling an interval-ordered task graph with uniform node weights to an arbitrary number of processors. However, the communication time among tasks of the parallel program is assumed zero [6]. Given these observations, the general scheduling problem cannot be solved in polynomial time unless $P = NP$ [7].Thus, a proper algorithm for the optimal solution in polynomial time is unlikely to exist.

The remainder of this paper is organized as follows; in section two, we review some related works. The third section contains task scheduling problem definition. In this section the input task graph, processor environment and objective of problem are explained. The Artificial Immune System is presented in section four. Discussion of our purposed algorithm is carried out in section five. The experimental results are illustrated and analyzed in section six. The last section provides conclusions and future works on this problem.

## 2. RELATED WORK

Methods in literature for solving static task scheduling are categorized into different classes based on the characteristics of both the decomposed tasks and the interconnected multiprocessor [6]. These methods can be classified into three categories in general: (1) deterministic based methods, that intuitively use some properties of a problem (heuristics) to solve it, (2) non-deterministic methods, which use random search techniques, namely metaheuristics, reach to solution, and (3) the hybrid of deterministic and non-deterministict methods.

The first category can be further classified in to three groups: list-scheduling heuristic algorithms, clustering heuristic algorithms and task duplication heuristic algorithms. List-scheduling methods are included of two classes, static list-scheduling and dynamic list-scheduling. In list-scheduling methods [8-10], at first, one priority list of tasks is created with respect to some heuristics. Then, one task is selected from the list and given to processors on the bases of some other heuristics (for example task is designated on a processor with minimum processor time) and the limitation of problem (preference conditions). In static list-scheduling algorithms, the initial priority list does not change until the algorithm finishes, but in dynamic List-scheduling algorithms, the priority of tasks that still are not allocated to one processor, is calculated. Furthermore, the arrangement of them is changed in priority list which leads to better results. The deterministic based methods due to the use of heuristics, reach the solution in better complexity time comparing to non-deterministic ones. However, the drawback of these methods is that they are unlikely to produce consistent results for a wide range of problems. The reason is that some heuristics are not persistent due to changing of properties of the problem. There are various list-scheduling heuristic studies for static task scheduling problem such as Modified Critical Path (MCP) [10], Mapping Heuristic (MH) [11] and Dynamic Critical Path (DCP) [12]. Clustering heuristic algorithms attempt to allocate a group of tasks with high communication data dependency, namely one cluster task, on the same processor even other processors are idle, Hence, with reducing interprocessor communication cost, the makespan of scheduling will be improved. In these methods, it is assumed that unbounded number of processors are available in the beginning and then be decreased to real number with merging cluster tasks on different processors [13]. Some examples of these methods are Dominant Sequence Clustering (DSC) [14], Linear Clustering method [15] and Clustering and Scheduling System (CASS) [13]. Task duplication heuristic algorithms allocate duplication of some tasks on more than one processor, redundantly, thereby the interprocessor communication cost will be reduced [16, 17].

The second category includes population-based algorithms such as PSO, Genetic, AIS [17-20] and also includes trajectory algorithms like Variable Neighbourhood Search (VNS), simulated annealing algorithms [21, 22]. These methods find the solution in longer time in comparison to the above mentioned methods. But they are applicable for wide range of problems due to their random searching nature.
In the third category, hybridization of two previous categories is proposed. Accordingly, the useful properties of each category are employed and combined to overcome the drawbacks. Some related methods in the literature are introduced [21]. We will use hybridization approach in this paper.

# 3. Problem Definition

Task scheduling problem involves an application decomposed into smaller tasks with precedence constrains and data communications among them, heterogeneous computing environment that tasks of the application must run on them, and the makespan objective that should be optimized.

## 4.1 Application

In this regard, we present the application by a Directed Acyclic Graph (DAG), because it can display precedence constrains and communications among tasks. The DAG is specified by (T, E, W) collection as:

- T, is the set of nodes of graph that shows tasks of the application. $T_i$ depicts the i-th task of graph.
- E, is the set of arrow edges of graph that shows the precedence of tasks so that one task cannot start before all its predecessors. $e_{ij}$ depicts the edge between $v_i$ and $T_j$. It also indicates that $v_i$ is the predecessor of $T_j$ and $T_j$ is the successor of $T_i$. As illustrated in figure 1.a, $T_2$ is the predecessor of $T_3$ and $T_4$, $T_3$ and $T_4$ are successors of $T_2$.
- W, is the set of labels on edges that shows data communication among tasks. $w_{ij}$ depicts the data value required $T_j$ from task $T_i$. This value is labeled on $e_{ij}$ edge, see figure 1.a.

The task without any predecessors is called entry task. On the other hand, the task without any successors is called an exit task. In this study, we consider that only one task is entry task. To reach this consideration, we add one task ($T_0$) to the T that has zero processor time on each processor, and put it as predecessor of all of the entry tasks of graph. We put one arrow edge between $T_0$ and each of entry tasks, and set the initial data communication value between it and each of them to zero. Therefore, we provide the status that only one task of graph, is the entry task. Figure 1.a represents an example of one application modeled by DAG G= (T, E, C), where T = {$T_1$, $T_2$, $T_3$, $T_4$, $T_5$}, E = {$e_{12}$, $e_{23}$, $e_{24}$, $e_{35}$, $e_{45}$}, W = {10, 5, 2, 2, 7}, $T_1$ is the entry task and $T_5$ is the exit task. Also we show the number of tasks with Tc.

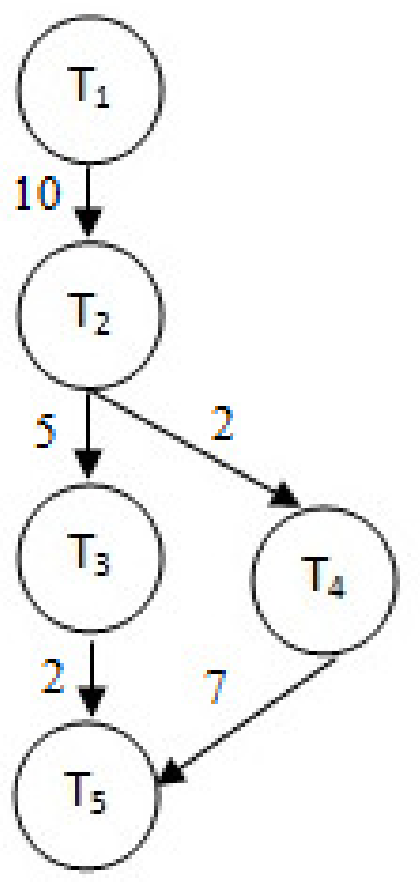

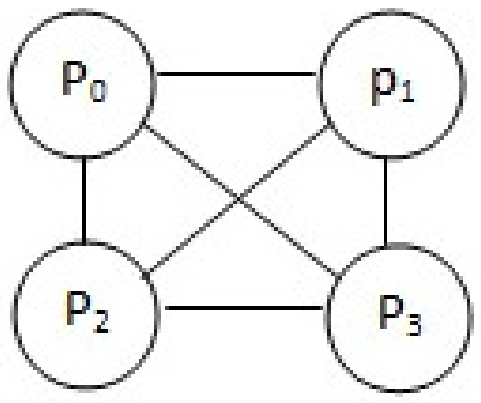

(b)

| | $P_0$ | $P_1$ | $P_2$ | $P_3$ |
|---|---|---|---|---|
| $T_1$ | 5 | 7 | 4 | 5 |
| $T_2$ | 5 | 4 | 4 | 6 |
| $T_3$ | 6 | 7 | 5 | 5 |
| $T_4$ | 7 | 5 | 5 | 4 |
| $T_5$ | 7 | 6 | 6 | 4 |

(c)

(a)

1. a) an example of DAG with 5 tasks, b) an example of fully connected network with 4 processors, c) the required processing time of each task on all processor.

Various type of DAGs have been introduced which indicate macro data-flow of parallel programs written in a SPMD style for distributed-memory systems to serve task scheduling problems, such as Fast Fourier Transformation (FFT) [23], Guess-Jordan (G-J) [24], Guess-elimination (G) [25], LU-decomposition [26] and Laplace equation solver [27]. Random Task Graph that produces random data flow graphs is also used in test beds [28]. In this paper, we assume that each task of DAG needs different processing time.

## 4.2 Environment

The environment in this problem is the Cluster Of Workstation (COW) which involves processing capability of each processors, bandwidth of communication links among them, and some other properties such as communication methods among processors, the probability of link/node failure, fully or not-fully connected processors. Environment in the primary studies is considered homogenous (a COW with the same processor and bandwidth of interprocessor communication links). However, the recent methods consider more real environment in which the COW has different processors and bandwidth of communication links (heterogeneous). In this study, we assume heterogeneous environment with fully connected processors, message-passing method for data communication and link and node failure. Assume, P is the set of processors, $p_i$ depicts the i-th processor and $P_c$ is the number of processors. Figure 1.b illustrates one fully connected COW with four processors $\{p_0, p_1, p_2, p_3\}$. Figure 1.c illustrates the processing time of each task on all the four heterogenous processors.

## 4.3 The objective of this paper

We consider the task scheduling problem with makespan objective and formally display it as function $F : T \rightarrow P$ , that maps tasks on processors. Let $V_i = \{T_j \in T \mid F(T_j) = p_i\}$ depict the set of all tasks that is allocated on processor i. So, the completion time of the processor i is the time that the last scheduled task on it be finished and given by;

$$C_i = AFT(T_j);\ \textit{where } T_j \in V_i \textit{ is the last scheduled task on } p_i. \quad (1)$$

Where $AFT(T_j)$ is the Actual Finish Time of $T_j$ defined as;

$$\mathrm{AFT(Tj)} = \{AST(Tj) + PT\ (Tj, F(Tj))\} \quad (2)$$

Where PT $(T_j, p_i)$ is the Processor Time of task $T_j$ on the processor $p_i$. Similarly, AST is the Actual Start Time that a task actually is started on this time. A task is ready when all its predecessor tasks are finished and its required data is met. In this time, the task can be started to execute. This time is called Earliest Start Time (EST). EST can formally be expressed as follows;

$$\mathrm{EST(Ti)} = Max\,\{\mathrm{AFT(Tj)} + \mathrm{wij} * \mathrm{ProcessorComm(F(Ti), F(Tj))} \mid \mathrm{Tj} \in \mathrm{predecessor(Ti)}\} \quad (3)$$

Where

$$\mathrm{ProcessorComm}(pi, pj) = \begin{cases} 0, & if\ i = j \\ rij, & if\ i \neq j \end{cases} \quad (4)$$

$r_{ij}$ is the processor communication cost between two processors i and j, the $AST(T_j)$ is the time that task is ready to processing and processor has not any other higher priority ready tasks to run. So $AST(T_j)$ can be given by:

$$AST(Tj) = \{EST(Tj) + waitingTime(Tj)\} \quad (5)$$

waitingTime($T_j$) is the time that the ready task $T_j$ is waiting for higher priority ready tasks on the processor to their processing are finished.
So, the makespan of the problem is calculated by;

$$Cmax = Max\{Ci\} \text{where i=0,1,…,p} \qquad (6)$$

Up to now, we have formulated the makespan. To formulate a function that provides a good decision about the better solutions, we used the fitness function as bellow;

$$fitness = \frac{F1 - Min_F1}{Max_F1 - Min_F1} \qquad (7)$$

Accordingly, we put F1 = Cmax from (6) (makespan) and The Min_F1 and Max_F1 are the minimum and maximum value of makespan index of the solutions respectively.

# 4. Artificial Immune System (AIS)

The natural immune system is one of the most important organs of our body that protect it from the infectious foreign elements (pathogens) and some abnormal behaviors of self elements of body. White blood cells, also called lymphocytes, are very important constituents of the immune system. These cells are created in the bone marrow, flow in the blood and lymph system, and exist in various lymphoid organs to do immunological functions. B and T cells constitute the main population of lymphocytes [29]. The surface of B-cells and T-cells are covered with receptors. When a pathogen is recognized by B-cell receptors, it is proliferated by cloning and differentiated. Some of these cloned cells are plasma cells, also known as antibody, that combat with antigens (or pathogens) and destroy them. Some others are memory cells that memorize the type of attack and lead to more speedy reactions of immune system when it faces on the same attacks later. Also, T-cells when recognized the pathogen, are proliferated by cloning. After the cloning, the affinity between antibodies and antigens is improved by the mutation of antibodies called as hypermutation. Hence the antibodies bind to antigens better and hence, combat with them better. The cloning, hypermutation and selection processes are the essential parts of the Clonal Selection Principle. Another important concept of immune system is the Immune Network (IN) theory introduced by Niels K. Jerne [30]. This theory tells that antibodies are also stimulated and recognized by each other. When an antibody recognizes another one, it is stimulated and cloned, whereas recognized antibody is suppressed. This cloning and suppression (immune-remove) lead the immune system to remain stable by avoiding inordinate antibodies is produced. The Immune Network Theory and Clonal Selection Principle are two main inspired concepts of immune system that make it proper for MOO problems. As the IN and Clonal Selection Principle cause to the diversity of population while provide the good search to find solutions simultaneously.

# 5. The proposed Method

The proposed method consist of two parts: (1) heuristic-based Artificial Immune System (figure 2), and (2) Single Neighbourhood Search. The SNS operation is done during scheduling.

## 5.1. Heuristic-based AIS algorithm

We start the AIS algorithm with the random population. This population improves in the various stages. Each individual named antibody, represent a candidate solution for the task scheduling problem. Antibodies are encoded as the strings of integers. Cell-indexes of the string depict task numbers and value of each cell of string represents the processor which task is allocated to it. Suppose there are $T_c$ tasks and $P_c$ processors, so the strings composed of $T_c$ cells and each cell of the string has the number between 0 and $P_c$-1. The example of two antibodies for one problem with 6 tasks and 4 processors is illustrated in figure 4. Each antibody has 6 cells which value of each cell is between 0 and 3.

After generating the initial random population, the order of task execution (as describe bellow) is determined for each antibody of population, and the makespan be calculated. Then, the clonal selection phase of the AIS is performed. The Clonal selection leads to widespread search around one point in the solution space and selection the best point (solution) from them, that it is tradeoff between exploration and exploitation.
After the Clonal selection, the immune-remove phase (immune network) of AIS is performed. This phase provides the elitism. The elitism and random population insertion of the algorithm in the step 2.f of figure 2 provide the diversity for the next population. The algorithm is iterated K times, where K is the input value of the algorithm, and eventually, the best antibody which is the best of the all population and the order of task execution, is returned as the answer.

Antibody(i):

$T_0$ $T_1$ $T_2$ $T_3$ $T_4$ $T_5$

| | | | | |
|---|---|---|---|---|
| 3 | 2 | 0 | 2 | 1 |

Antibody(j):

$T_0$ $T_1$ $T_2$ $T_3$ $T_4$ $T_5$

| | | | | |
|---|---|---|---|---|
| 3 | 1 | 0 | 3 | 2 |

Figure4. The example of two antibodies.

### 5.1.1. Clonal Selection

The Clonal selection phase of AIS is shown in steps 2.a.v until 2.a.ix of the algorithm in figure 2. First, the antibody, which is undergoing the cloning process, $C_{clone}$ times is cloned, where $C_{clone}$ is a positive integer value and the input parameter of the algorithm. Then, each clone is mutated, that is two random cells of the clone are selected and their values are swapped. The order of task execution of mutated clone is determined, and its makespan be calculated. After this, the minimum and maximum makespan among the antibody and mutated clones are found. Now, according to (7) the fitness of the antibody and each mutated clone will be calculated, and that solution which has the best fitness is replaced to the antibody in the population.

### 5.1.2. Immune-remove

The B number of the antibodies of the population that have better fitness is selected, where the B is a positive integer value and the input parameter of algorithm. When an antibody is selected, the affinity between the selected antibody and all of the other antibodies in the population is calculated. If the affinity between one antibody of the population and the selected antibody is lower than Aff , it is removed from the population, because this similar antibody is ignored in the next searching for the best. The Aff value is a non-negative integer and the input parameter of the algorithm. The affinity between two antibodies i and j is calculated as;

$$affinity(i,j) = \frac{distance(i,j)}{Tc} \qquad (8)$$

The distance between one pair of antibodies is computed according to Hamming distance, so that two antibodies are compared cell by cell with each other and if the k-th cell of $antibody_i$ is different from the k-th cell of $antibody_j$ where k=0,…, $T_c$-1, then the distance increases a unit. For example in the figure 5, the distance between $antibody_i$ and $antibody_j$ which is displayed by distance (i, j) is equal to 3.

Inputs; K: an integer; Popsize: an integer; Clone: an integer; B: an integer; Aff: a positive real value lower than1;
Outputs; An antibody with order of tasks execution of it.
Algorithm:
1. Initialize a random population of antibodies
2. Repeat the following K times
   a. For each antibody in population
      i. Map the tasks on processors on the basis of encoded antibody.
      ii. Schedule the order of task execution for the antibody.
      iii. Calculate the makespan of antibody.
      iv. Calculate the reliability of antibody.
      v. Clone the antibody $C_{clone}$ times and for each Clone
         1. Perform the mutation operator.
         2. Map the tasks on processors on the basis of mutated Clone.
         3. Determine the order of task execution for the Clone.
         4. Calculate the makespan of Clone.
      vi. Determine the maximum and minimum makespan among Clones and the antibody.
      vii. Calculate the fitness of antibody and Clones base on (7).
      viii. Select one that has the best fitness as the antibody of population.
   b. Repeat the following, B times
      i. Select the best antibody from the population.
      ii. Calculate the distance between the best antibody and the each other antibodies of population.
      iii. If distance between best antibody and that antibody is lower than Aff then remove it from the population.
      iv. Save best antibody in memory.
   c. Insert all the memory's antibodies in the next population.
   d. Add random antibodies in next population until the size of it gets the Popsize.
3. Return best antibody and order of task execution of it as output.

Figure2. The hybrid heuristic-based AIS algorithm.

#### 5.1.3. Heuristic approach for determining the order of task execution of one solution

The detailed heuristic procedure of determining the order of task execution is:

Input; The solution string; Task graph;
Output; The order of task execution of the solution;
Algorithm:
1. Calculate the rank of each task of the encoded solution.
2. For each processor in the parallel of other processors do
   a. Create an empty priority queue.
   b. The task which has no predecessor, adds in the queue of processor base on solution string.
   c. If the queue is not empty, perform the SNS and take one task from the first of priority queue and execute it.
   d. For all tasks of this processor which all the predecessors of them are executed, insert them in the proper place of queue.
   e. Repeat the steps 2.c and 2.d until all tasks which are allocated on this processor are executed.

Figure 3. The heuristic procedure of schedule the order of task execution.

The rank of each task in one solution is calculated on the basis of sum of b-level value of successors of it as:

$$rank(Ti) = \sum_{j} b - level(Tj)\,; Tj \in \text{ successor(Ti)} \qquad (9)$$

The b-level($T_i$) in one solution is the length of a longest path from $T_i$ to an exit task node of the task graph. The b-level is accounted as:

Inputs; The task graph with precedent constrains and data flow communications(W); The processing time of tasks on processors; The solution string;
Output The b-level of each task of graph for the solution;
Algorithm:

1. Construct a list of tasks in reversed topological order. Call it RevTopList.
2. For each $T_i$ in the RevTopList
    a. max = 0.
    b. If $T_i$ has some successors
        i. max = Max {$w_{ij}$ * ProcessorComm ($T_i$, $T_j$) + b-level($T_j$) | $T_j \in$ successor of $T_i$}.
        ii. b-level($T_i$) = max + PT($T_i$, F($T_i$)).
    c. else
        i. b-level($T_i$) = PT($T_i$, F($T_i$)).

Figure 4. the algorithm of calculation b-level of each task respect to the solution.

After calculation the b-level of tasks, one priority queue of tasks that all predecessors of them are executed will be constructed for each processor. For all processors, the following actions are performed simultaneously. First, the task which has no any predecessors ($T_0$) adds in the queue of the proper processor according to solution string. When one task is processed, the data streams from it, is flowed to the successor tasks. Each time some tasks which allocated on this processor for insertion in the queue are prepared, they are inserted in to the proper places in the priority queue. The insertion does on the basis of the rank of tasks and the time of processor. The tasks in the priority queue are ordered on the basis of their ranks. The impact of giving the time of processor in proper place is that there may some tasks of queue was processed in this time and so the new entry tasks in the queue cannot be inserted before those processed task, although the rank of new task is further from them. This insertion of tasks is continued until all tasks insert in their queues.
Before the selection of one task from the queue, the SNS is performed on the queue and then one task is selected.

## 5.2. Single Neighbourhood Search (SNS)

There may some processors be idle as waiting for data flow from other processors. Therefore, there are not any ready tasks for processing. However, sometimes processors have some ready tasks for processing, but the order of task execution on one processor leads to the processor that is being idle. In this situation, some tasks with lower ranks could be processed in the idle times. This drawback can be recognized during the scheduling and can be eliminated by SNS. However, the idle time of processors cannot be eliminated completely by using SNS, because it is considered that the tasks processing are non-preemptive, but SNS tries to reduce the problem as possible as. In one solution, the priority queue of each processor determines the order of task execution of it. In each priority queue, if the difference between AFT of task that are processing and AST of the next task for processing in the queue is bigger than or equal to the processing time of each other waiting tasks in the queue, the task which has higher priority from these waiting tasks be selected to run as next task.

## 6. Experiments

To test the effectiveness of our algorithm, we obtain the solutions for the applications of some realistic problem task graphs comparing to some other recent methods. In addition, we have compared these methods with HAIS to show the effect of SNS. We use the parameter settings shown in table 1 in all runs.

Table 1. Parameter settings used in the experiment.

| **parameter** | **value** |
|---|---|
| Number of iterations(K) | 100 |
| Population size (Popsize) | 400 |
| Number of clones(Clone) | 50 |
| Selection rate (B) | 0.25 |

At first, we consider 18 nodes Gaussian Elimination test graph in [12] which is shown in figure 7. Figure 8 shows the example of task graph with 9 tasks. The Gantt chart of one found solution of our purposed algorithm for this graph is illustrated in figure 9.Table 2 shows that the results of our purposed algorithm for graph in figure 7. These results predicate that proposed method is better than MPC, DSC, MD, DCP, HAIS, and acts as well as the PMC-GA [31] and [32] methods. These results also show the impact of SNS in purposed algorithm. Furthermore this table shows that our algorithm finds the better result than MPC, DSC, MD, DCP, HAIS, and are as well as PMC-GA and [32] for graph in figure 8. The results in table 3 also indicate the effectiveness of the proposed method for solving this problem.

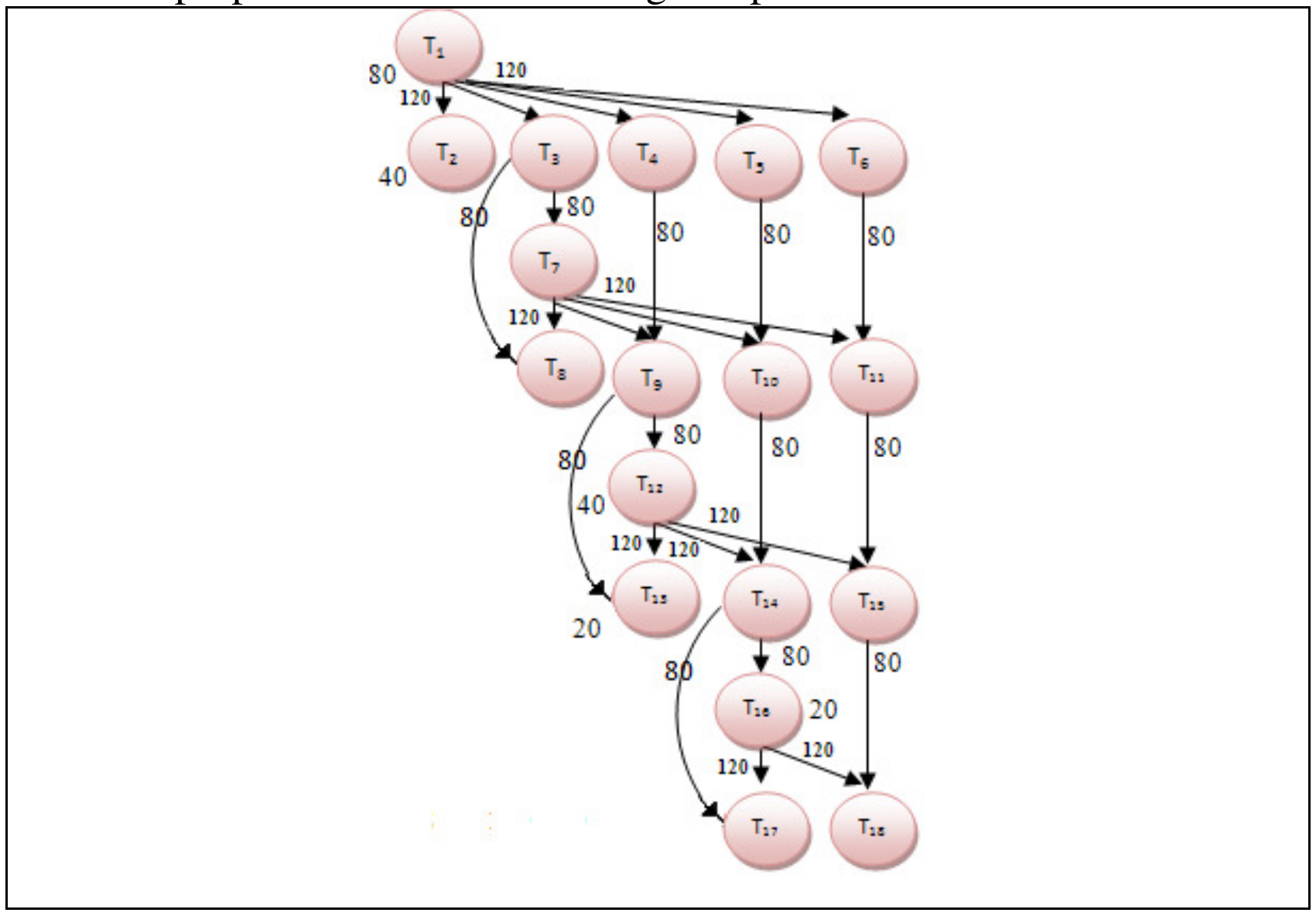

Figure 7. 18 nodes Gaussian Elimination test graph[12].

Table 2. Comparative results of purposed algorithm with others for 18 Gaussian Elimination test graph.

| Algorithm | MCP | DSC | MD | DCP | PMC-GA | [32] | HAIS | Purposed algorithm |
|---|---|---|---|---|---|---|---|---|
| No. processors | 4 | 6 | 3 | 3 | 2 | 2 | 2 | 2 |
| Finish Time for fig. 7 | 520 | 460 | 460 | 440 | 440 | 440 | 450 | 440 |
| No. processors | 3 | 4 | 2 | 2 | 2 | 2 | 2 | 2 |
| Finish Time for fig. 8 | 29 | 27 | 32 | 32 | 23 | 21 | 21 | 21 |

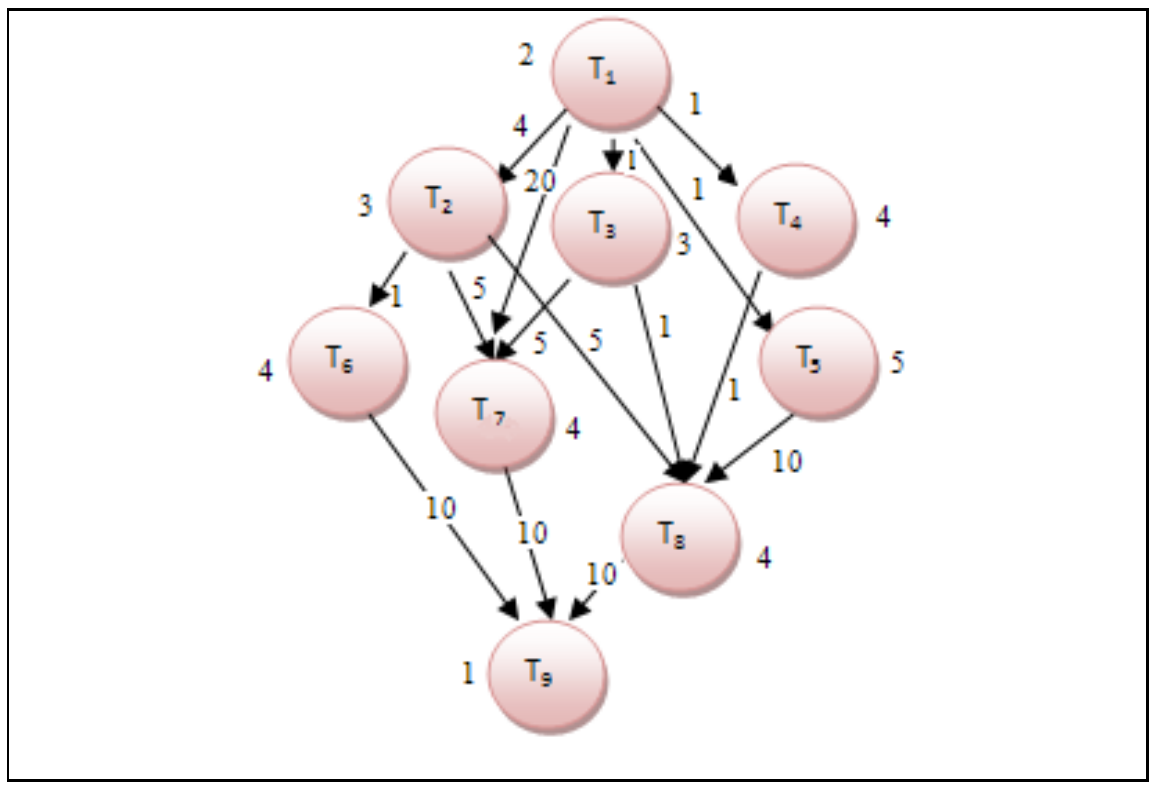

Figure 8. Example of task graph with 9 tasks [33].

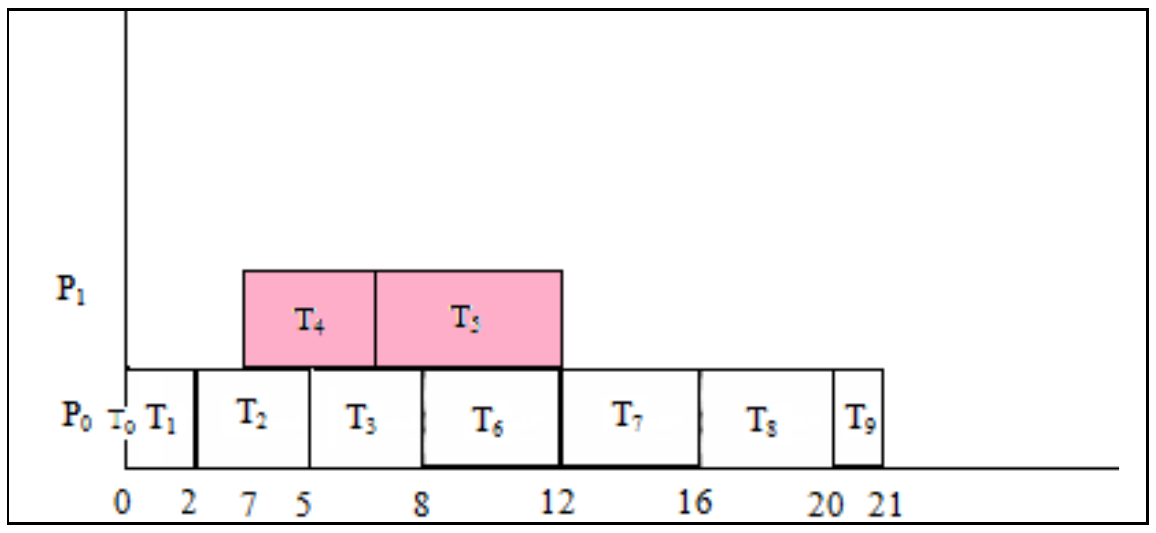

Figure 9. Gantt chart of purposed algorithms.

## 7. Conclusions

In this paper we purposed the hybrid heuristic AIS algorithm for task scheduling problem in a COW. This algorithm has applied the two main operators of AIS in compose of one ranking method. Then a local search method that is SNS for utilization results is done. The results show that this algorithm can effectively solve this problem with various graphs. Our futures work is to apply our algorithm to solve multiobjective scheduling problem, since the AIS algorithm is capable for solving multiobjective problems. Furthermore, the finishing time of scheduling algorithms is serious. Some cells of immune system memorized the improved antibodies (solutions) for rapid defense in next attack of the same antigen. This property of immune system (memorization) can also use in scheduling problem for time saving to find solution.